\documentclass[twocolumn,showpacs,preprintnumbers,prl]{revtex4}
\usepackage{graphicx}
\usepackage{subfigure}
\usepackage{dcolumn}
\usepackage{bm}
\usepackage{amssymb}
\usepackage{amsmath}
\usepackage{epsfig}
\usepackage{graphics}
\usepackage[dvips]{color}

\newcommand{\imsizeA}{0.95\columnwidth}

\begin{document}

\title{Spin-orbit interaction in symmetric wells with two subbands}
\author{Esmerindo Bernardes$^1$}
\author{John Schliemann$^{2,3}$}
\author{Minchul Lee$^{3}$}
\author{J. Carlos Egues$^{1,3,4}$}
\author{Daniel Loss$^{3,4}$}
\affiliation{$^1$ Instituto de F\'{\i}sica de S\~{a}o Carlos, Universidade de S\~{a}o
Paulo, 13560-970 S\~{a}o Carlos, S\~{a}o Paulo, Brazil}
\affiliation{$^{2}$Institute for Theoretical Physics, University of Regensburg, D-93040
Regensburg, Germany}
\affiliation{$^{3}$Department of Physics and Astronomy, University of Basel, CH-4056
Basel, Switzerland}
\affiliation{$^4$Kavli Institute for Theoretical Physics, University of California, Santa
Barbara, California 93106, USA}
\date{\today }

\begin{abstract}
We investigate the spin-orbit (s-o) interaction in two-dimensional
electron gases (2DEGs) in quantum wells with two subbands. From the
$8\times 8$ Kane model, we derive a new inter-subband-induced s-o
term which resembles the functional form of the Rashba s-o -- but is
non-zero even in \emph{symmetric} structures. This follows from the
distinct parity of the confined states (even/odd) which obliterates
the need for asymmetric potentials. We self-consistently calculate
the new s-o coupling strength for realistic wells and find it
comparable to the usual Rashba constant. Our new s-o term gives rise
to a non-zero ballistic spin-Hall conductivity, which changes sign
as a function of the Fermi energy ($\varepsilon_F$), and can induce
an unusual \emph{zitterbewegung} with cycloidal trajectories
\textit{without} magnetic fields.

\end{abstract}


\maketitle

The rapidly developing field of spintronics has generated a great
deal of interest in s-o coupling in semiconductor nanostructures
\cite{overview}. For an n-doped zincblende semiconductor quantum
well with only the lowest subband occupied, i.e. in a strictly 2D
situation, there are two main contributions to the interaction of
the spin and orbital degrees of freedom of electrons. One
contribution is the Dresselhaus term, which results from the lack of
inversion symmetry of the underlying zinc-blende lattice
\cite{Dresselhaus55} and is to lowest order linear in the crystal
momentum \cite{Dya-bas}. This linearity is shared by the other
contribution known as the Rashba term \cite{Rashba60}, which is due
to structural inversion asymmetry and can be tuned by an electric
gate across the well \cite{nitta}. These two contributions can lead
to an interesting interplay in spintronic systems \cite
{Schliemann03a}.

In this Letter we consider yet another type of electronic s-o
coupling which, as we show, occurs in III-V (or II-VI) zinc-blende
semiconductor quantum wells with more than one subband. We derive a
new inter-subband-induced s-o interaction which resembles that of
the ordinary Rashba model; however, in contrast to the latter, ours
is non-zero even in \emph{symmetric} structures, Fig. 1. We self
consistently determine the strength of this new s-o coupling for
realistic single and double wells and find it comparable to the
Rashba constant, Figs. 2(a)-(b). We have investigated the spin Hall
effect and the dynamics of spin-polarized electrons due to this new
s-o term. We find (i) a non-zero ballistic spin Hall conductivity
which changes sign as a function of $\varepsilon_F$ and (ii) an
unusual \emph{zitterbewegung} \cite{zitter1} with cycloidal
trajectories without magnetic fields, Fig. 3. As derived below, for
a symmetric well with two subbands our $4\times 4$ electron
Hamiltonian is
\begin{eqnarray}
\mathcal{H} &=&\left( \frac{\vec{p}^{2}}{2m^{\ast }}+\epsilon _{+}\right)
\mathbf{1}\otimes \mathbf{1}-\epsilon _{-}\tau ^{z}\otimes \mathbf{1}  \notag \\
&+&\frac{\eta }{\hbar }\tau ^{x}\otimes \left( p_{x}\sigma ^{y}-p_{y}\sigma
^{x}\right) \,,  \label{twobandham}
\end{eqnarray}
where $m^{\ast }$ is the effective mass, $\epsilon _{\pm }=\left(
\varepsilon _{o}\pm \varepsilon _{e}\right) /2$, $\varepsilon _{e}$ and $%
\varepsilon _{o}$ are quantized energies of the lowest (even) and first
excited (odd) subbands (corresponding to eigenstates $|e\rangle $ and $%
|o\rangle $), respectively, measured from the bottom of the quantum well, $%
\tau ^{x,y,z}$ denote the Pauli matrices describing the subband (or
\begin{figure}[t]
\begin{center}
\epsfig{file=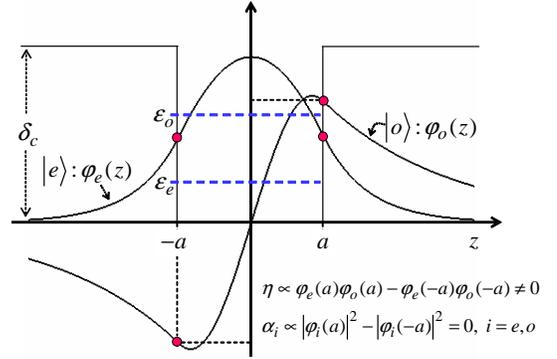, width=0.40\textwidth}
\end{center}
\par
\caption{Square well with its ground-state
${\protect\varphi_{e}(z)}$ and first excited-state $\protect\varphi
_{o}(z)$ wavefunctions. The new
inter-subband-induced s-o coupling $\protect\eta $ in Eq. (\protect\ref{eta}%
) is non-zero even in \textit{symmetric} wells due to the distinct parities
of $\protect\varphi _{e}(z)$ (even) and $\protect\varphi _{o}(z)$ (odd),
which yield a non-vanishing matrix element for the derivative of the
symmetric potential.}
\label{fg1}
\end{figure}
pseudospin) degree of freedom, and $\sigma ^{x,y,z}$ are Pauli matrices
referring to the electron spin. The new \emph{inter-subband-induced} s-o
coupling $\eta $ is
\begin{eqnarray}
\eta &=&-\left( \frac{1}{E_{g}^{2}}-\frac{1}{\left( E_{g}+\Delta \right) ^{2}%
}\right) \frac{P^{2}}{3}\langle e|\partial _{z}V(z)|o\rangle \,  \notag \\
&&+\left( \frac{\delta _{v}}{E_{g}^{2}}-\frac{\delta _{\Delta
}}{\left( E_{g}+\Delta \right) ^{2}}\right) \frac{P^{2}}{3}\langle
e|\partial _{z}h(z)|o\rangle ,\,  \label{eta}
\end{eqnarray}%
where $E_{g}$ and $\Delta $ are the fundamental and split-off band gaps in
the well region \cite{Vurgaftman01}; $P$ is the Kane matrix element \cite%
{kane}. The parameters $\delta _{v}$ and $\delta _{\Delta }$ denote
valence band offsets between the well and the barrier regions
\cite{deltas}, $V(z)$ \ is the Hartree-type contribution to the
electron potential and $h(z)$ is the structural quantum-well profile
\cite{lassnig}. Note that $\eta $ can be varied via external gates,
Fig. 2. Next we outline the derivation of $\mathcal{H}$ in Eq. (\ref
{twobandham}).

\emph{Kane Hamiltonian. }We start from the usual $8\times 8$ Kane
Hamiltonian describing the $s$-type conduction and the $p$-type valence
bands around the $\Gamma $ point \cite{Trebin79},
\begin{equation}
\mathcal{H}_{8\times 8}=\left(
\begin{array}{cc}
H_{c} & H_{cv} \\
H_{vc} & H_{v}%
\end{array}%
\right) ,  \label{kane}
\end{equation}%
where $H_{c}$ is a $2\times 2$ diagonal matrix with elements $
p^{2}/2m_{0}+V_{c}(\vec{r})$, $m_{0}$ is the bare electron mass,
$H_{v}$ is a $6\times 6$ diagonal matrix with elements $p^{2}/2m_{0}+V_{v}(\vec{r}%
)-E_{g}$ for the heavy- and light-hole bands and $p^{2}/2m_{0}+V_{\Delta }(%
\vec{r})-E_{g}-\Delta $ for the split-off band, $V_{i}(\vec{r})$ [$i=c$, $v$%
, $\Delta $] denote arbitrary potentials (see below), and $H_{cv}=$
$\left( H_{vc}\right) ^{\dag }$ is
\begin{equation}
H_{cv}=\left(
\begin{array}{cccccc}
\frac{-\kappa _{+}}{\sqrt{2}} & \sqrt{\frac{2}{3}}\kappa _{z} & \frac{\kappa
_{-}}{\sqrt{6}} & 0 & \frac{-\kappa _{z}}{\sqrt{3}} & \frac{-\kappa _{-}}{%
\sqrt{3}} \\
0 & \frac{-\kappa _{+}}{\sqrt{6}} & \sqrt{\frac{2}{3}}\kappa _{z} & \frac{%
\kappa _{-}}{\sqrt{2}} & \frac{-\kappa _{+}}{\sqrt{3}} & \frac{\kappa _{z}}{%
\sqrt{3}}%
\end{array}%
\right) \,,  \label{Hcv}
\end{equation}%
where $\vec{\kappa}=P\vec{k}$, $\vec{k}=\vec{p}/\hbar $ is the electron wave
vector, $k_{\pm }=k_{x}\pm ik_{y}$, and $P=-i\hbar \langle S|p_{x}|X\rangle
/m_{0}$ parameterizes the conduction-to-valence band coupling; $|S\rangle $
and $|X\rangle $ are the usual periodic Bloch functions at the $\Gamma $
point.

\emph{Effective electron Hamiltonian: folding down. }The Kane Hamiltonian (%
\ref{kane}) acts on an eight-component spinor $\Psi ^{\dag }=\left(
\begin{array}{cc}
\psi _{c} & \psi _{v}%
\end{array}%
\right) ^{\dag }$ in which the last six components $\psi _{v}$ represent
valence-band states. \ By eliminating the hole components from the Schr\"{o}%
dinger equation $\mathcal{H}_{8\times 8}\Psi =\varepsilon \Psi $, where $%
\varepsilon $ is the eigenenergy, we can fold down this $8\times 8$ equation
into a $2\times 2$ effective equation for the conduction-band states only: ${%
\mathcal{H(\varepsilon )}}\tilde{\psi}_{c}=[H_{c}+H_{cv}\left( \varepsilon
-H_{v}\right) ^{-1}H_{vc}]\tilde{\psi}_{c}$, $\tilde{\psi}_{c}$ is a
renormalized conduction-electron spinor.

\emph{s-o in symmetric wells. }Applying the above procedure to a
quantum well, defined by the confining potentials \cite{lassnig}
$V_{i}(\vec{r})\rightarrow V_{i}(z)=V(z)+\delta _{i}h(z)$, $i=c$,
$v$, $\Delta $, we find
\begin{eqnarray}
{\mathcal{H(\varepsilon )}} &=&H_{QW}+\frac{P^{2}}{3\hbar ^{2}}p_{-}\left[
\eta _{1}^{-1}+\eta _{2}^{-1},p_{z}\right] ,  \label{genefham} \\
H_{QW} &=&p_{+}\frac{1}{2m^{\ast }(z,\varepsilon )}p_{-}+p_{z}\frac{1}{%
2m^{\ast }(z,\varepsilon )}p_{z}+V_{c}(z),  \label{Hqw}
\end{eqnarray}%
where $1/m^{\ast }(z,\varepsilon )=\frac{2P^{2}}{3\hbar ^{2}}\left( \frac{2}{%
\eta _{1}}+\frac{1}{\eta _{2}}\right) +\frac{1}{m_{0}}$, $\ \eta
_{1}=\varepsilon -(p^{2}/2m_{0}+V(z)-\delta _{v}h(z)-E_{g})$ and $\eta
_{2}=\varepsilon -(p^{2}/2m_{0}+V(z)-\delta _{\Delta }h(z)-E_{g}-\Delta )$.
Equation (\ref{genefham}) \ describes an electron in a quantum well ($H_{QW}$
term) with spin orbit interaction (last term) \cite{darwin}. The
kinetic-energy operators above are complicated due to the position- and
energy-dependent effective mass $m^{\ast }(z,\varepsilon )$. Since $E_{g}$
and $E_{g}+\Delta $ are the largest energy scales in our system, we can
simplify (\ref{genefham}) and (\ref{Hqw}) by expanding $1/\eta _{1}$ and $%
1/\eta _{2}$ in the form $1/\eta _{1}=$ $E_{g}^{-1}\{1-[\varepsilon
-p^{2}/2m_{0}-V(z)+\delta _{v}h(z)]/E_{g}+...\}$ and $1/\eta _{2}=$
$\left( E_{g}+\Delta \right) ^{-1}\{1-[\varepsilon
-p^{2}/2m_{0}-V(z)+\delta _{\Delta }h(z)]/\left( E_{g}+\Delta
\right) +...\}$. \ To zeroth order $\eta _{1}=E_{g}$, $\eta
_{2}=E_{g}+\Delta $ and $H_{QW}=p_{\Vert }^{2}/2m^{\ast
}+p_{z}^{2}/2m^{\ast }+V_{c}(z)$ with (a constant effective mass)
$1/m^{\ast
}=\frac{2P^{2}}{3\hbar ^{2}}\left( \frac{2}{E_{g}}+\frac{1}{E_{g}+\Delta }%
\right) +\frac{1}{m_{0}}$ \cite{eff-mass}. Since the s-o operator $\left[
\eta _{1}^{-1}+\eta _{2}^{-1},p_{z}\right] \rightarrow \partial _{z}(1/\eta
_{1})+\partial _{z}(1/\eta _{2})$, we need to keep the first-order terms in
the expansions of $\eta _{1}^{-1}$ and $\eta _{2}^{-1}$ which yield the
leading non-zero contribution to the s-o term in (\ref{genefham}). We find $%
\left[ \eta _{1}^{-1}+\eta _{2}^{-1},p_{z}\right] =[1/E_{g}^{2}-$ $1/\left(
E_{g}+\Delta \right) ^{2}]\partial _{z}V(z)-[\delta _{v}/E_{g}^{2}-$ $\delta
_{\Delta }/\left( E_{g}+\Delta \right) ^{2}]\partial _{z}h(z)$. Finally, we
project this s-o operator into the \emph{two} lowest (spin-degenerate)
eigenstates $|i\rangle _{\sigma _{z}}=|\vec{k}_{\Vert }i\rangle |\sigma
_{z}\rangle $, $\langle \vec{r}|\vec{k}_{\Vert }i\rangle =\exp (i\vec{k}%
_{\Vert }\cdot \vec{r}_{\Vert })\varphi _{i}(z)$, $i=e,o$ and
$\sigma _{z}=\uparrow ,\downarrow $, of the \emph{symmetric} well
($H_{QW}$), Fig. 1. This directly leads to the ${\mathcal{H}}$ in
(\ref{twobandham}) with the s-o coupling $\eta $ (\ref{eta})
\cite{sym}. Note that this new s-o interaction is non-zero even in
symmetric wells as $\eta$ arises from the coupling between the
ground state (even) and the first excited state (odd), Eq.
(\ref{eta}). We can generalize ${\mathcal{H}}$ to include the Rashba
$\alpha$ and the linearized Dresselhaus $\beta$ s-o couplings. Next
we determine the magnitude of $\eta$ (and $\alpha$, $\beta$) for
realistic quantum wells with two subbands.

\emph{Self-consistent calculation of the s-o couplings.} We consider
modulation-doped quantum wells similar to those experimentally
investigated in Ref. \cite{koga}. Our wells, however, have
\textit{two} occupied subbands. Similarly to Ref. \cite{koga}, we
study cases with constant chemical potentials \cite{n-fixed}. By
self-consistently solving Poisson and Schr\"odinger's equations we
determine the energy levels $\varepsilon_e$, $\varepsilon_o$ and the
confined wave functions $\varphi_{i}(z)$, $i=e,o$ of the wells. We
then calculate (i) $\eta$ via Eq. (\ref{eta}), (ii) $\alpha_i$ from
equations similar to Eq. (\ref{eta}) for each subband, and (iii)
$\beta_i=\gamma_c\langle i| k_z^2|i \rangle$, $\gamma_c$ is the bulk
Dresselhaus s-o parameter \cite{jancu}. The structural symmetry of
the wells and their charge densities can be changed via a gate
potential $V_b$.

Our calculated s-o couplings $\eta$, $\alpha_i$, and $\beta_i$,
$i=e,o$, for an InAlAs/InGaAs/InAlAs \textit{single} quantum well
(``sample 3'' in \cite{koga}) are all comparable in magnitude, Fig.
2(a). Note that our $\alpha_e/\beta_e$ ratio is consistent with the
experimental one in Ref. \cite{gani-prb}. In addition, $\alpha_e$
vs. $V_b$ here agrees well with the experimental data in Fig. 3
(``triangle up'') of Ref. \cite{koga} (see also Fig. 4 in
\cite{koga1}). Our $\alpha_i$  \cite{hu} and $\beta_i$ are also
consistent with those of Ref. \cite{saikin}. Note that for the
single well studied here $\eta$ does not vary appreciably with the
gate $V_b$, similarly to $\beta_i$ and as opposed to $\alpha_i$.
\begin{figure}[h]
\begin{center}
\epsfig{file=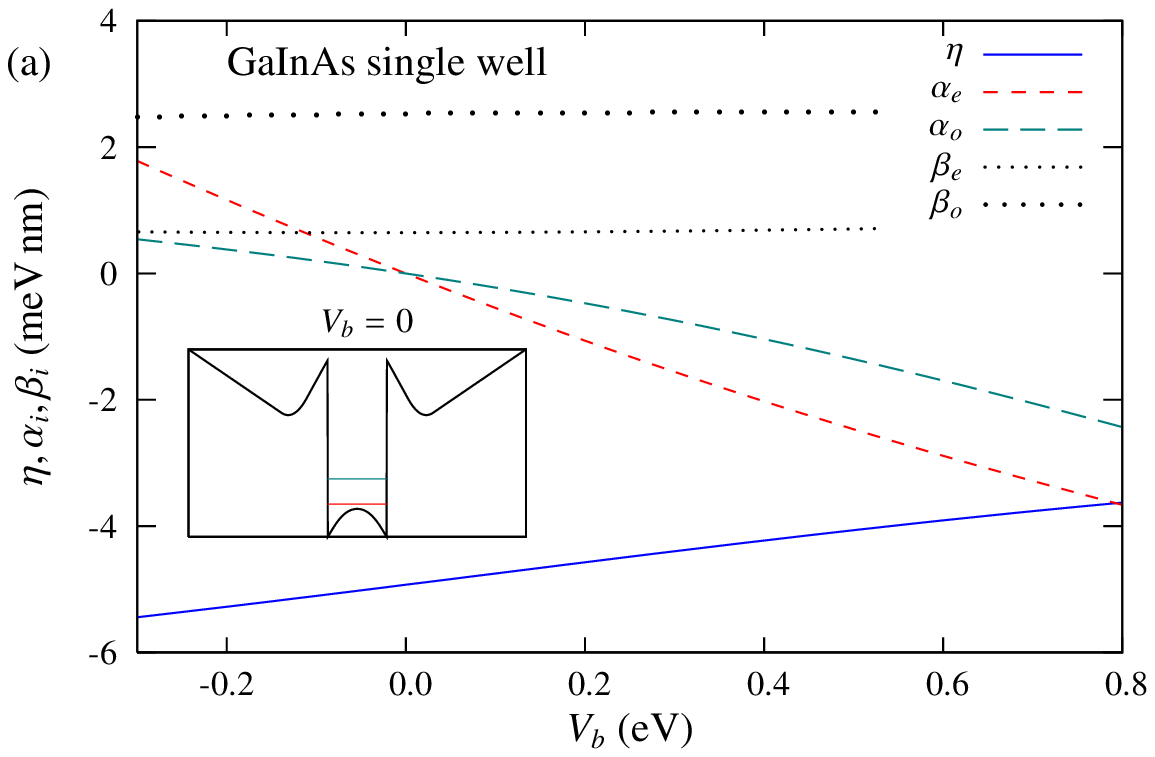, width=0.4\textwidth}
\epsfig{file=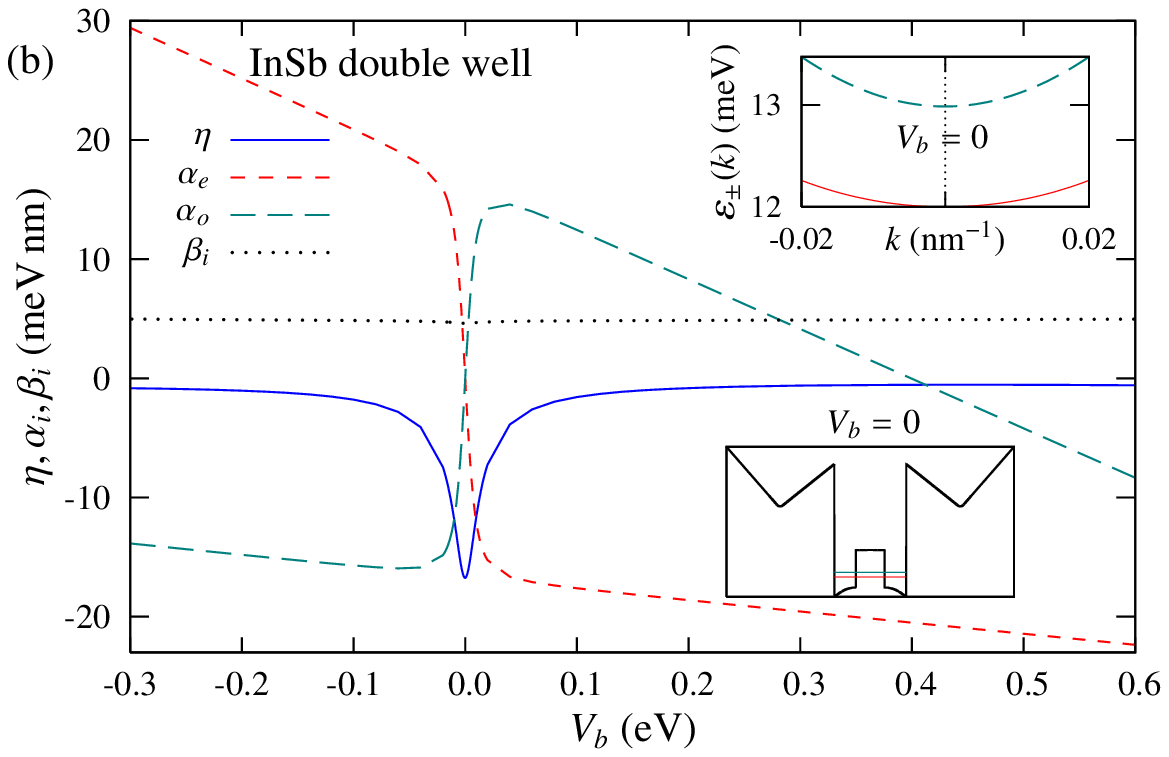, width=0.4\textwidth}
\end{center}
  \par
  \caption{\label{fig:GaInAs} Calculated s-o coupling strengths
  as a function of the external gate $V_b$ for
  realistic wells. (a) For the single GaInAs \cite{koga,koga1} well
  studied, the inter-subband coupling $\eta$ is larger than
  the Dresselhaus $\beta_i$ and the Rashba $\alpha_i$ constants
  ($i=e,o$). Note that $|\alpha_e|\ge|\alpha_o|$ and both change sign
  across $V_b=0$ (in constrast to $\beta_i$ and $\eta$). (b) For the
  InSb double well considered, $\eta$ shows a ``resonant behavior''
  about $V_b=0$ [symmetric configuration, lower-left inset in (b)]. This occurs
  because the subband splitting $\varepsilon_o - \varepsilon_e$
  reaches a minimum at $V_b=0$ and the double-well wave functions
  are very similar (though of distinct parities) for $V_b \sim 0$.
  This also makes $\alpha_e \sim - \alpha_o$ around $V_b=0$.
  Upper-right inset in (b): energy dispersions
  $\varepsilon _{\pm }(\vec{k})$ [Eq. (\ref{eigenvalues})]
  of the symmetric double well.}
\end{figure}

For a \textit{double-well} structure, on the other hand, we find
that $\eta$ has a ``resonant behavior'', changing by about an order
of magnitude as $V_b$ is swept across $V_b=0$, Fig. 2(b) [this may
have a dramatic effect on Shubnikov-de Haas measurements]. $V_b=0$
corresponds to a fully symmetric double well. In contrast to the
single-well case, $\alpha_e$ and $\alpha_o$ have opposite signs and
undergo abrupt changes in magnitudes near $V_b=0$ [Fig. 2(b), dashed
lines]. Similarly to the single well case, $\beta_e$ and $\beta_o$
are also essentially constant for a double well [Fig. 2(b), dotted
lines]. A detailed account of our results will be presented
elsewhere. Having established that the new s-o coupling $\eta$ is
sizable, in what follows we focus on a fully symmetric well to
investigate physical effects arising solely from $\eta$.

\emph{Fully symmetric case: eigensolutions.} Let us consider a two
subband well (single or double) described by the Hamiltonian
$\mathcal{H}$ in (\ref{twobandham}) (we assume a negligible
Dresselhaus term \cite{gani-prb}). In the basis $\left\{ |e\rangle
_{\uparrow },|o\rangle _{\downarrow },|o\rangle _{\uparrow
},|e\rangle _{\downarrow }\right\} $ $\mathcal{H}$ becomes
\begin{equation}
\tilde{{\mathcal{H}}}=\left(
\begin{array}{cccc}
\frac{\hslash ^{2}k^{2}}{2m^{\ast }}+\varepsilon _{e} & -i\eta k_{-} & 0 & 0
\\
i\eta k_{+} & \frac{\hslash ^{2}k^{2}}{2m^{\ast }}+\varepsilon _{o} & 0 & 0
\\
0 & 0 & \frac{\hslash ^{2}k^{2}}{2m^{\ast }}+\varepsilon _{o} & -i\eta k_{-}
\\
0 & 0 & i\eta k_{+} & \frac{\hslash ^{2}k^{2}}{2m^{\ast }}+\varepsilon _{e}%
\end{array}
\right) .  \label{Hblock}
\end{equation}%
Both the upper-left ($U$) and lower-right ($L$) blocks of $\tilde{{\mathcal{H%
}}}$ have eigenvalues
\begin{equation}
\varepsilon _{\pm }(\vec{k})=\epsilon _{k}\pm \hbar \Omega ,
\label{eigenvalues}
\end{equation}
with $\epsilon _{k}=\frac{\hbar ^{2}k^{2}}{2m^{\ast}}+\epsilon
_{+}$, $(\hbar \Omega )^{2}=\eta ^{2}k^{2}+\epsilon _{-}^{2}$, and
eigenvectors
%\begin{equation}
%|\psi _{1}\rangle _{+}^{U} =\sin \left( \theta /2\right) |e\rangle
%_{\uparrow }+\cos \left( \theta /2\right) e^{i\phi } |o\rangle
%_{\downarrow } \text{,}  \label{eigenvector1}
%\end{equation}
\begin{eqnarray}
|\psi _{1}\rangle _{+}^{U} &=&\sin \left( \theta /2\right) |e\rangle
_{\uparrow }+\cos \left( \theta /2\right) e^{i\phi } |o\rangle _{\downarrow }%
\text{,}  \label{eigenvector1} \\
\text{ }|\psi _{2}\rangle _{+}^{L} &=&\cos \left( \theta /2\right)
|o\rangle _{\uparrow }+\sin \left( \theta /2\right)e^{i\phi }
|e\rangle _{\downarrow }, \\ \label{eigenvector2}
%\end{eqnarray}%
%\begin{eqnarray}
|\psi _{3}\rangle _{-}^{U} &=&\cos \left( \theta /2\right) |e\rangle
_{\uparrow }-\sin \left( \theta /2\right) e^{i\phi }|o\rangle _{\downarrow },
\label{eigenvector3} \\
|\psi _{4}\rangle _{-}^{L} &=&\sin \left( \theta /2\right) |o\rangle
_{\uparrow }-\cos \left( \theta /2\right) e^{i\phi }|e\rangle
_{\downarrow }. \label{eigenvector4}
\end{eqnarray}
Here, $e^{i\phi }=(-k_{y}+ik_{x})/k$, $\cos (\theta
)=1/\sqrt{1+\left( \eta k/\epsilon _{-}\right) ^{2}}$, and
$\vec{k}=k(\sin \phi ,-\cos \phi $) (here we drop the
\textquotedblleft $\Vert $\textquotedblright\ in $\vec{k}_{\Vert
}$). For $\eta k<<2\epsilon _{-}$ we can expand $\varepsilon _{\pm
}(\vec{k})$ in (\ref{eigenvalues}) and define effective masses
$m_{\pm }^{\ast }=m^{\ast }/[1\pm 2\varepsilon _{so}/\epsilon
_{-}]$, where $\varepsilon _{so}=\eta ^{2}m^{\ast }/2\hbar ^{2}$ is
the energy scale of the new s-o coupling. For the double well of
Fig. 2(b), $m^\ast_{-}$ is reduced by $\sim 5 \%$ compared to the
bulk value $m^\ast$. This could be measured via, e.g.,
cyclotron-resonance experiments \cite{cr}.

\emph{Novel Zitterbewegung.} The dynamics of electron wave packets
in wells with s-o interaction exhibit an oscillatory motion
\cite{zitter1} -- the \emph{zitterbewegung}. For our new s-o
interaction, a wave packet $|\chi \rangle $ moves according to $
\langle \chi |\vec{r}_{H}(t)|\chi \rangle $ where
$\vec{r}_{H}(t)=U^{\dagger }\vec{r}U$ is the position operator
in the Heisenberg picture [$U=\exp (-i%
\mathcal{H}t/\hslash )$] with components
\begin{eqnarray}
x_{H}(t) &=&\mathbf{1}\otimes \mathbf{1}x(0)+\mathbf{1}\otimes \mathbf{1}%
\frac{p_{x}}{m^{\ast }}t+\frac{\eta }{\hbar }t\tau ^{x}\otimes \sigma ^{y}
\notag \\
&-&\frac{\eta }{2(\hbar \Omega )^{2}}\left[ \epsilon _{-}\tau ^{y}\otimes
\sigma ^{y}+\frac{\eta }{\hbar }p_{y}\mathbf{1}\otimes \sigma ^{z}\right]
%\times
%\notag \\
%&&
\left(\cos \left( 2\Omega t\right) -1\right)  \notag \\
&+&\frac{\eta }{2(\hbar \Omega )^{3}}\Bigl[\epsilon _{-}^{2}\tau
^{x}\otimes \sigma ^{y}+\epsilon _{-}\frac{\eta }{\hbar }\tau
^{z}\otimes
\mathbf{1}p_{x}  \notag \\
&+&\left( \frac{\eta }{\hbar }\right) ^{2}p_{y}\tau ^{x}\otimes
\left( p_{x}\sigma ^{y}+p_{y}\sigma ^{x}\right) \Bigr]
\left(\sin\left(2\Omega t\right)-2\Omega t\right), \label{xH}
\end{eqnarray}%
and $y_{H}(t)$, obtained from Eq. (\ref{xH}) via the replacements $%
(p_{x},\sigma ^{x})\mapsto (p_{y},\sigma ^{y})$, $(p_{y},\sigma
^{y})\mapsto (-p_{x},-\sigma ^{x})$ [i.e. a $\pi /2$ rotation about
the $ z $-axis]. Similar expressions can be derived for the spin
components $\sigma^i_{H}(t)$, $i=x,y,z$ \cite{bsel-pasps}.

For simplicity, we evaluate the expectation value of
$\vec{r}_{H}(t)$ for planes waves (\textquotedblleft wide wave
packets\textquotedblright ). For a spin-up electron injected into
the lowest subband along the $y$-axis with (group) velocity
$\vec{v}_{g}=(\hbar k_{0y}/m^{\ast })\hat{y}$ we find
\begin{eqnarray}
\langle x_{H}(t)\rangle &=&\frac{\eta ^{2}k_{0y}}{2(\hbar \Omega )^{2}}%
\left( 1-\cos \left( 2\Omega t\right) \right),  \label{xh} \\
\langle y_{H}(t)\rangle &=&\frac{\hbar k_{0y}}{m^{\ast
}}t+\frac{\eta ^{2}k_{0y}\epsilon _{-}}{2(\hbar \Omega )^{3}}\left(
\sin \left( 2\Omega t\right) -2\Omega t\right),  \label{yh}
\end{eqnarray}%
assuming $x(0)=y(0)=0.$ Equations (\ref{xh}) and (\ref{yh}) show
that cycloidal motion is possible in our system. This differs
qualitatively from the Rashba s-o \emph{zitterbewegung} which is
always perpendicular to the initial $\vec{v}_{g}$.
\begin{figure}[h]
\centering
{\resizebox{\imsizeA}{!}{\includegraphics{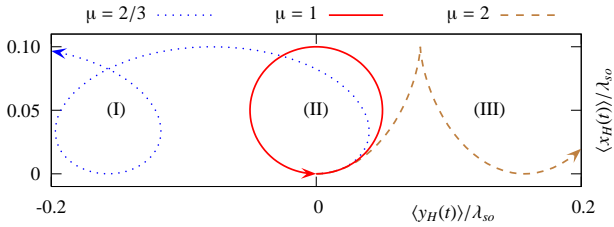}}}\newline
\par
\caption{Zitterbewegung due to the s-o coupling $ \protect\eta $ for
distinct ratios $\mu =\epsilon _{-}/2\varepsilon _{so}$. Note the
peculiar trajectories with the \textit{forward} injected electrons
moving \textit{backward} (I) and even in a \textit{closed} path
(II). This follows from the s-o induced change in the curvature of
the bands which renormalizes the effective masses. Here we use
$\protect\lambda _{so}k_{0y}=\protect\mu/10$,
$\protect\lambda_{so}^{-1}=m^{\ast }\protect \eta /\hbar .$}
\label{fig:pwypanel}
\end{figure}

Figure 3 shows trajectories for three distinct $\vec{v}_{g}=(\hbar
k_{0y}/m^{\ast })\hat{y}$ -- all with $k_{0y}>0$. We find motion
\emph{opposite to} and\emph{\ }along the $y$-axis (orbits I and III,
resp.) and even a closed path (II). To understand this behavior we
note that for $\eta k_{0y}<<\varepsilon _{o}-\varepsilon _{e}$ the
linear-in-$t$ terms in $\langle y_{H}(t)\rangle $ can be recast into
$\hbar k_{0y}t/m_{-}^{\ast }$ $\Rightarrow$ the injected wave moves
with the renormalized velocity $v_{g}^{\ast }=\hbar
k_{0y}/m_{-}^{\ast }$. Hence, for orbit I: $\mu <1\Rightarrow
m_{-}^{\ast }<0$ \cite{sena} and $v_{g}^{\ast }<0$; for orbit II:
$\mu =1\Rightarrow m_{-}^{\ast }\rightarrow \infty \ $and
$v_{g}^{\ast }=0$; and  for orbit III: $\mu \,>1\Rightarrow
m_{-}^{\ast }>0$ and $v_{g}^{\ast }>0$. Though remarkable, we stress
that the orbits I and II occur for \textit{unusual} parameters
(e.g., $\varepsilon _{F}<\varepsilon _{so}/10$). However, these
orbits do show that our s-o Hamiltonian has a physical mechanism
allowing for cyclotronic motion without magnetic fields.

\emph{Spin Hall conductivity $\sigma_{xy}^z$.} The spin Hall effect
is a convenient probe for s-o effects in wells \cite{she-exp}. We
have calculated $\sigma_{xy}^z$ (``clean limit'') in the presence of
an external magnetic field $B$ by following the approach of Rashba
\cite{rashba2004}, which allows us to properly account for both the
intra- and inter-branch contributions in the Kubo formula
\cite{sinova}. Here we focus on the $B\rightarrow 0$ limit where we
find $\sigma^z_{xy}=0$ for $\varepsilon_F
> \varepsilon_{o}$ (two subbands occupied) and
\begin{align}
  \sigma^z_{xy} & =
  \frac{e}{8\pi}
  \left[
    \frac{1}{\kappa_1}\left(\frac{1}{\kappa_3} - 2\right)
    +
    \frac{\kappa_2 + \kappa_1/2}{2\kappa_3^3}
  \right]\,
\end{align}
for $\varepsilon_{e} < \varepsilon_F < \varepsilon_{o}$ (upper
subband empty), where $\kappa_1 = 2\varepsilon_{so}/\epsilon_{-}$,
$\kappa_2 = (\varepsilon_F - \epsilon_{+})/2\epsilon_{-}$, and
$\kappa_3 = \sqrt{\kappa_1^2/4 + \kappa_1\kappa_2 + 1/4}$. Note that
$\sigma^z_{xy}$ is non-zero and non-universal in this range, shows a
discontinuity at $\varepsilon_F=\varepsilon_o$, and changes sign as
a function of $\varepsilon_F$.  Details of our calculation of
$\sigma^z_{xy}$ and a thorough discussion will be presented
elsewhere \cite{she-calc}. Here we just note that measurements of
$\sigma^z_{xy}$ (vs. $\varepsilon_F$) in symmetric two-subband wells
offer a possibility to probe our new  s-o interaction
\cite{she-exp}. Note that the Rashba and the linearized Dresselhaus
spin Hall conductivities are identically zero in the DC
limit\cite{rashba2004,sinova1}.

We have introduced an inter-subband-induced s-o interaction in
quantum wells with two subbands. The corresponding s-o coupling
$\eta $ (whose magnitude is similar to Rashba coupling) is non-zero
even in symmetric wells. This new s-o interaction gives rise to a
non-zero spin Hall conductivity, renormalizes the bulk mass by $\sim
5\%$ (measurable via cyclotron resonance \cite{cr}) in double wells
and can induce a cycloidal \emph{zitterbewegung}. Weak
antilocalization \cite{koga1,wal} should offer another possibility
to measure $\eta$.

We thank S. Erlingsson,\ D. S. Saraga, D. Bulaev, J. Lehmann, M.
Duckheim, L. Viveiros, G. J. Ferreira, R. Calsaverini, and E. Rashba
for useful discussions. This work was supported by the Swiss NSF,
the NCCR Nanoscience, DARPA, ONR, CNPq, FAPESP, DFG via SFB 689, and
USA NSF PHY99-07949.


\begin{thebibliography}{99}
\bibitem{eff-mass} We treat $m^{\ast }$ as a parameter in our model (the
Kane effective mass neglects corrections from higher-lying bands).
\bibitem{sym} In Eq. (\ref{twobandham}) we assume that $\langle e|\partial
_{z}V|o\rangle $ and (in turn) $\eta $ are real. If $\eta =|\eta |e^{i\delta
}$, we have to replace $\tau ^{x}$ by $\cos (\delta )\tau ^{x}-\sin (\delta
)\tau ^{y}$ in (\ref{twobandham}). The freedom in fixing $\delta $ has no
physical consequences since a change in $\delta $ can be compensated by
applying the phase factor $\exp (i\delta \tau ^{z})$ to the basis states: $%
[\cos (\delta )\tau ^{x}-\sin (\delta )\tau ^{y}]\exp (i\delta \tau
^{z})=\tau ^{x}$. Thus, ${\mathcal{H}}$ enjoys a U(1) gauge symmetry
corresponding to a rotation about the $z$-direction in the
$\vec{\tau}$ space.
\end{thebibliography}
\end{document}